\documentclass[12pt,twocolumn,twoside]{pnas-new}

\setcitestyle{authoryear,open={(},close={)}}

\usepackage{balance} 
\usepackage{fancyvrb}

\templatetype{pnasresearcharticle} 

\usepackage{times}
\usepackage{urcsbiblio}
\usepackage{latexsym}
\usepackage{multirow}

\usepackage{graphicx}

\usepackage{xspace}
\usepackage{amssymb,amsmath,epsfig}

\usepackage{amsthm}
\usepackage{algorithm}
\usepackage[noend]{algpseudocode}

\newcommand{\Tstrut}{\rule{0pt}{2.6ex}}         

\newcommand{\nts}{\ensuremath{\text{\it nt}}\xspace}
\newcommand{\pairs}{\ensuremath{\mathrm{pairs}}\xspace}
\newcommand{\unpaired}{\ensuremath{\mathrm{unpaired}}\xspace}

\newcommand{\ppv}{\ensuremath{\text{PPV}}\xspace}

\newcommand{\notes}[1]{}

 \theoremstyle{definition}
 
\theoremstyle{plain}

\newcommand{\ith}[1]{\ensuremath{i^{{th}}}}

\newcount\permx
\newcount\permy
\def\permdot#1#2{
\permx=#1 \advance\permx by-1
\permy=#2 \advance\permy by-1
\psframe[fillcolor=black, fillstyle=solid]
(\permx,\permy)(#1, #2)
}

\newcommand{\x}[1]{\ensuremath{x_{#1}}\xspace}

\newcommand{\yhat}{\ensuremath{\hat{y}}\xspace}
\newcommand{\ystar}{\ensuremath{y^*}\xspace}

\definecolor{mylightgray}{rgb}{0.85,0.85,0.85}
\newcommand{\myboxmath}[1]{\ensuremath{\fcolorbox{white}{mylightgray}{\!\!\ensuremath{#1}\!\!}}\xspace}
\definecolor{lightblue}{rgb}{0.8,0.9,1}
\newcommand{\myboxmathblue}[1]{\ensuremath{\fcolorbox{white}{lightblue}{\!\!\ensuremath{#1}\!\!}}\xspace}

\newcommand{\vecy}{\ensuremath{\mathbf{y}}\xspace}

\newcommand{\smallnt}[1]{\ensuremath{_{\mbox{\tiny PP}}}\xspace}

\newcommand{\pseudocode}{Algorithm}
\floatname{algorithm}{\pseudocode}

\iffalse

\else

\fi

\newcommand{\smallurl}[1]{{\scriptsize \url{#1}}}

\newcommand{\push}{\ensuremath{\mathsf{push}}\xspace}

\newcommand{\pop}{\ensuremath{\mathsf{pop}}\xspace}

\newcommand{\nskip}{\ensuremath{\mathsf{skip}}\xspace}

\newcommand{\rnastructure}{{RNAstructure}\xspace}

\newcommand{\md}{\ensuremath{\text{\tt .}}}

\newcommand{\mla}{\ensuremath{\text{\tt (}}}
\newcommand{\mra}{\ensuremath{\text{\tt )}}}
\newcommand{\bml}{\ensuremath{\text{\tt{\textbf (}}}}

\newcommand{\ml}{\mla}
\newcommand{\mr}{\mra}

\newcommand{\mq}{\ensuremath{\text{\tt ?}}}

\newcommand{\linearfold}{{LinearFold}\xspace}

\newcommand{\contrafold}{{CONTRAfold}\xspace}

\newcommand{\viennarna}{{ViennaRNA}\xspace}
\newcommand{\viennarnafold}{{Vienna RNAfold}\xspace}
\newcommand{\rnafold}{{RNAfold}\xspace}

\newcommand{\myurl}[1]{\href{{#1}}{\tt\scriptsize {#1}}} 

\newcommand{\probknot}{ProbKnot\xspace}
\newcommand{\ipknot}{IPknot\xspace}
\newcommand{\threshknot}{ThreshKnot\xspace}
\newcommand{\pkiss}{pKiss\xspace}

\newcommand{\linearpartition}{LinearPartition\xspace}

\newcommand{\crossing}{\ensuremath{\mathrm{crossing}}\xspace}
\newcommand{\PPV}{\ensuremath{\mathrm{\text{\sc ppv}}}\xspace}
\newcommand{\Sens}{\ensuremath{\mathrm{\text{sens}}}\xspace}
\newcommand{\ppvcrossing}{\ensuremath{\PPV_\crossing}\xspace}
\newcommand{\senscrossing}{\ensuremath{\Sens_\crossing}\xspace}
\newcommand{\Fscore}{\ensuremath{\mathrm{F}}\xspace}

\onecolumn 

\title{\threshknot: Thresholded \probknot for Improved RNA Secondary Structure Prediction}

\author[a]{Liang Zhang}
\author[b]{He Zhang}
\author[c]{David H.~Mathews}
\author[a,b,$\clubsuit$]{Liang Huang}

\affil[a]{School of Electrical Engineering~\& Computer Science,
  Oregon State University, Corvallis, OR}
\affil[b]{Baidu Research, Sunnyvale, CA}
\affil[c]{Department of Biochemistry \& Biophysics, Center for RNA Biology, and Department of Biostatistics \& Computational Biology, University of Rochester Medical Center, Rochester, NY}
\leadauthor{L.~Zhang}

\twocolumn  

\correspondingauthor{
\vspace{-0.7cm}
  \textsuperscript{$\clubsuit$}
  Corresponding author: {liang.huang.sh@gmail.com}. 
}

\begin{abstract}
RNA structure prediction is a challenging problem, especially with pseudoknots. 
Recently, there has been a shift from the classical minimum free energy-based methods (MFE)
to partition function-based ones that assemble structures using base-pairing probabilities.
Two examples of the latter group are the popular maximum expected accuracy (MEA) method
and the \probknot method.
\probknot is a fast heuristic that pairs nucleotides that are reciprocally most probable pairing partners, and unlike MEA, can also predict structures with pseudoknots.
However, \probknot's full potential has been largely overlooked. 
In particular, when introduced, it did not have an MEA-like hyperparameter
that can balance between positive predictive value (PPV) and sensitivity.
We show that a simple thresholded version of \probknot, which we call {\em \threshknot}, 
leads to more accurate overall predictions by filtering out unlikely pairs whose probabilities fall under a given threshold.
We also show that on three widely-used folding engines (RNAstructure, Vienna RNAfold, and CONTRAfold),
\threshknot always outperforms the much more involved MEA algorithm in
(1) its higher structure prediction accuracy,
(2) its capability to predict pseudoknots, and
(3) its faster runtime and easier implementation.
This suggests that \threshknot should replace MEA as the default partition function-based structure prediction algorithm.
\threshknot is already available in the widely used \rnastructure software package version 6.2 (released November 27, 2019):
\myurl{https://rna.urmc.rochester.edu/RNAstructure.html}
  \vspace{-1cm}
\end{abstract}

\dates{}

\begin{document}

\verticaladjustment{-2pt}

\maketitle
\thispagestyle{firststyle}
\ifthenelse{\boolean{shortarticle}}{\ifthenelse{\boolean{singlecolumn}}{\abscontentformatted}{\abscontent}}{}



\section{Introduction}
\label{sec:intro}
RNAs are involved in multiple processes,
including catalysis, guiding RNA modification, and post-transcriptional gene regulation \cite{doudna+cech:2002,bachellerie+:2002,storz+gottesman:2006,wu+belasco:2008,serganov+nudler:2013,karijolich+:2015}.
Often, RNA function is highly related to structure.
However, structure determine techniques, such as Cryo-Electron Microscopy (Cry-EM) \cite{ognjenovic+:2019},
X-ray crystallography \cite{zhang+erre-DAmare:2014}
or Nuclear Magnetic Resonance (NMR) \cite{zhang+keane:2019},
though reliable and accurate,
are slow and costly.
Therefore, fast and accurate computational prediction of RNA structure is useful and desired.
Because tertiary structure modeling
is challenging \cite{miao+:2017},
many studies focus on predicting the secondary structure,
i.e.,
the double helices formed by base pairing of self-complementary nucleotides (A-U, G-C, G-U base pairs) 
\cite{tinoco+bustamante:1999}.
The secondary structure is well-defined, 
provides detailed information to help understand the structure-function relationship,
and is a basis to predict full tertiary structure \cite{parisien+:2008,seetin+:2011,nawrocki+eddy:2013}.

Most algorithms for RNA secondary structure prediction can be divided
into two categories, the classical ones computing a single structure with the minimum free energy (MFE) \cite{nussinov+jacobson:1980,zuker+stiegler:1981},
and the more recent ones based on the partition function, which is the sum of all equilibrium constants for all possible structures and is the normalization for estimating marginal probabilities of base pairs and motifs \cite{mccaskill:1990}.
Generally speaking, there is a trend to shift from the former (MFE-based) methods to the latter (partition function-based) ones
for many reasons, including
(1) the overall accuracy of partition function-based methods is generally higher than that of MFE-based \cite{do+:2006,lu+:2009,hajiaghayi+:2012},
(2) instead of predicting a single structure as in MFE, the partition function captures the whole ensemble of conformations
and an RNA molecule (e.g., mRNAs) can be many different conformations at equilibrium \cite{tafer+:2008,lu+mathews:2008,long+:2007,cordero+das:2015,lai+:2018},
(3) we can also induce the base-pairing probabilities from the partition function,
and 
(4) as a by-product, heuristic algorithms can use the partition function to predict pseudoknots\footnote{
A pseudoknot involves at least two pairs $(i,j)$ and $(k,l)$ such that $i<k<j<l$. \label{footnote:crossing}}
\cite{bellaousov+mathews:2010,sato+:2011}. 


There are two typical (and widely used)
examples of partition function-based prediction algorithms.
The first is maximum expected accuracy (MEA) \cite{do+:2006,knudsen+hein:2003},
which predicts the structure \vecy that 
maximizes the sum of the base-paired  and single-stranded probabilities ($p_{i,j}$'s and $q_j$'s, respectively):
\[
  2\gamma\sum_{(i,j)\in \pairs(\vecy)}p_{i,j} + \sum_{j\in \unpaired(\vecy)}q_j
\]
where $\gamma$ is a  hyperparameter that balances the positive predictive value (PPV; a.k.a.~precision) and sensitivity (a.k.a.~recall) of the output structure.
The other one is \probknot \cite{bellaousov+mathews:2010},
which builds structure of mutually maximal probability pairing partners.
Both use base-pairing probabilities to assemble the output structure,
but the former requires another $O(n^3)$-time dynamic program for the assembly,
while the latter is a simpler heuristic method that only needs $O(n^2)$ time.
More importantly, \probknot can predict pseudoknots while MEA cannot.

However, the full potential of \probknot has not been fully exploited. 
In particular, unlike MEA, \probknot lacks a hyperparameter to balance PPV and sensitivity. 
To address this problem, we present \threshknot (short for \underline{Thresh}olded Prob\underline{Knot}),
which adds a probability threshold $\theta$ to disallow any pair whose probability falls below $\theta$.
Therefore, a smaller value of $\theta$ encourages \threshknot to predict more base pairs, and a higher one makes it more selective.
By tuning $\theta$, we can balance 
the PPV (the fraction of predicted pairs in the accepted structure) and 
sensitivity (the fraction of accepted pairs predicted). 


Simple as it is,
we show that \threshknot leads to more accurate overall predictions, 
and with three widely-used folding engines (\rnastructure \cite{reuter+:2010}, \viennarnafold \cite{lorenz+:2011}, and \contrafold \cite{do+:2006}),
\threshknot always outperforms the much more involved MEA algorithm 
in all three aspects: (1) it can achieve better overall prediction accuracy 
than MEA,
(2) it can predict pseudoknots that MEA can not,
(3) it is much simpler to implement and runs much faster.
This suggests that \threshknot should replace MEA as the default partition function-based structure prediction algorithm.




\section{Results}
\label{sec:result}
\begin{figure} 
    \centering
      \includegraphics[width=0.5\textwidth]{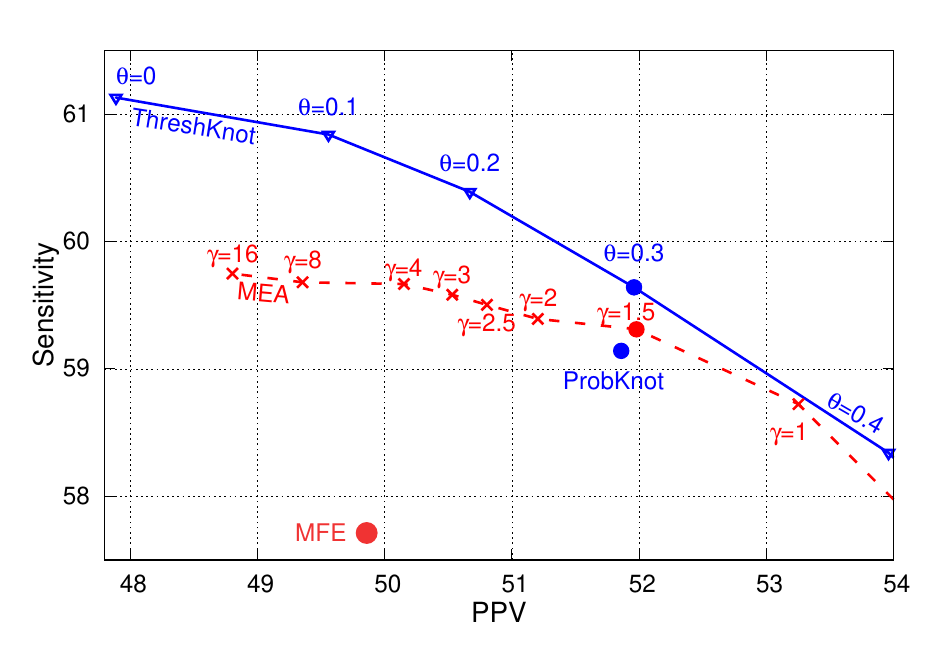}
\vspace{-.8cm}
  \caption{Comparison between minimum free energy structure prediction (MFE), maximum expected accuracy (MEA), \probknot, and \threshknot on \rnastructure 
    using the ArchiveII dataset of sequences with well-determined secondary structures. \threshknot has better PPV and Sensitivity than all other methods.
    For MEA and \threshknot, calculations were run as a function of the hyperparameter, $\gamma$ and $\theta$, respectively.  Between points, lines are draw.  \probknot and MFE have no tuning parameter and therefore are single points on these plots.
    \probknot filters out helices of two or less base pairs, while \threshknot only filters out single-pair helices.
  \label{fig:main_text_curves} }
\end{figure}

\subsection{\threshknot Algorithm}
\threshknot, like \probknot,
outputs the secondary structure 
made of ``most probable base pairs''.
i.e., pairs $(i,j)$
whose probability $p_{i,j}$ is the highest among ``competing pairs'',
i.e., $p_{i,j} \ge p_{i,k}$ for all $k$
and  $p_{i,j} \ge p_{l,j}$ for all $l$. 
But in addition to that, \threshknot 
also rules out any pair whose probability falls below $\theta$,
i.e.,
it returns the set of  pairs
\[
   \{ (i,j) \mid p_{i,j} = \max_k p_{i,k} = \max_l p_{l,j} \text{ and } p_{i,j} \geq \theta \}.
   \]

   To keep it simple, unlike \probknot
   which removes helices composed of two or less stacked pairs,
   \threshknot only removes single-pair helices.\footnote{
   The \threshknot results of not removing any helices are almost identical to those removing single-pair helices.}

\subsection{Overall Prediction Accuracy}
Below we show \threshknot results using the base-pairing matrices generated by RNAstructure;
see the Supplementary Information for the results of \threshknot on \contrafold and \viennarnafold. 
Figure~\ref{fig:main_text_curves} compares \threshknot with MEA, MFE, and \probknot. 
We choose $\theta$ = 0, 0.1, 0.2, 0.3, 0.4, 0.5, and 0.6 for \threshknot, 
and $\gamma$ = 0.5, 1, 1.5, 2, 2.5, 3, 4, 8, and 16 for MEA. 
We evaluate the overall prediction accuracies across all families, 
reporting both PPV and sensitivity. 

\begin{table}[]
\hspace{-0.5cm}
\resizebox{0.52\textwidth}{!}{
\setlength{\tabcolsep}{3pt}
\begin{tabular}{@{}c|c|l|r|r|r|r@{}}
\multicolumn{2}{c|}{} &        &\multicolumn{2}{c|}{\em overall}  &\multicolumn{2}{c}{\em pseudoknot}\\
\multicolumn{2}{c|}{} & time complexity & \multicolumn{1}{c}{PPV} & sens. & \multicolumn{1}{c}{PPV} & sens.\\
\hline
\multirow{4}{*}{\rotatebox{90}{\rnastructure\!\!}} & MFE & $O(n^3)$ & 49.86 & 57.71 & - & - \Tstrut \\
& MEA & $\myboxmath{O(n^3)} + \myboxmathblue{O(n^3)}$ & 51.98 &	59.31 & - & -\\
& \probknot & $\myboxmath{O(n^3)} + \myboxmathblue{O(n^2)}$ & 51.86 &	59.14 & 7.04 & 2.59\\
& \!\!\threshknot\!\!\! & $\myboxmath{O(n^3)} + \myboxmathblue{O(n^2)}$ & 51.96 & {\bf 59.64} & 7.62 & 2.85\\[0.05cm]
\hline
\multicolumn{2}{c|}{\ipknot} & $\myboxmath{O(n^3)} + \myboxmathblue{O(n^2)\! +\! \text{ILP}}$& {\bf 60.22} & 51.46 & {\bf 16.16} & 8.60 \\
\multicolumn{2}{c|}{\pkiss} &$O(n^4)$& 44.32 & 51.03 &  9.72	& {\bf 15.29}
\end{tabular}
}
\smallskip
 \caption{Accuracy of \threshknot.
The gray-shaded \myboxmath{O(n^3)} denotes the time to compute the partition function and base-pairing probabilities,
and \myboxmathblue{\text{ light blue shades }} denote the time for
post-processing steps based on those probabilities.
\myboxmathblue{\text{ILP}} denotes the time to solve the integer linear program,
which is  NP-complete in the worst case but very fast in practice.
See the Methods section for the definitions of pseudoknot PPV (\ppvcrossing) and Sensitivity (\senscrossing).
\label{tab:accuracy_table}
\vspace{-0.5cm}
 }
\end{table}

\begin{figure*}[!h]
\centering
  \begin{tabular}{ll}
\hspace{-0.1cm} {\bf A} & \hspace{-0.3cm} {\bf B}
\\[-0.5cm]
\hspace{-0.5cm}
    \includegraphics[width=0.52\textwidth]{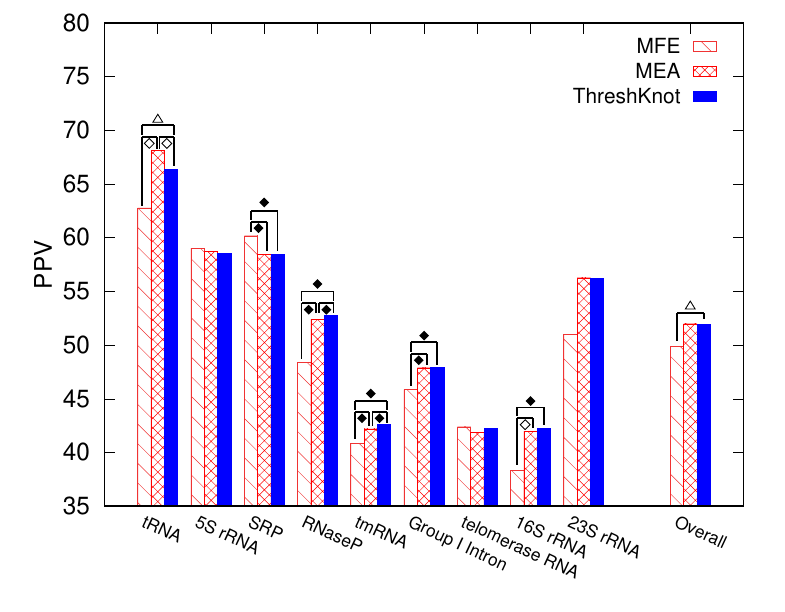}
&
\hspace{-0.7cm}
    \includegraphics[width=0.52\textwidth]{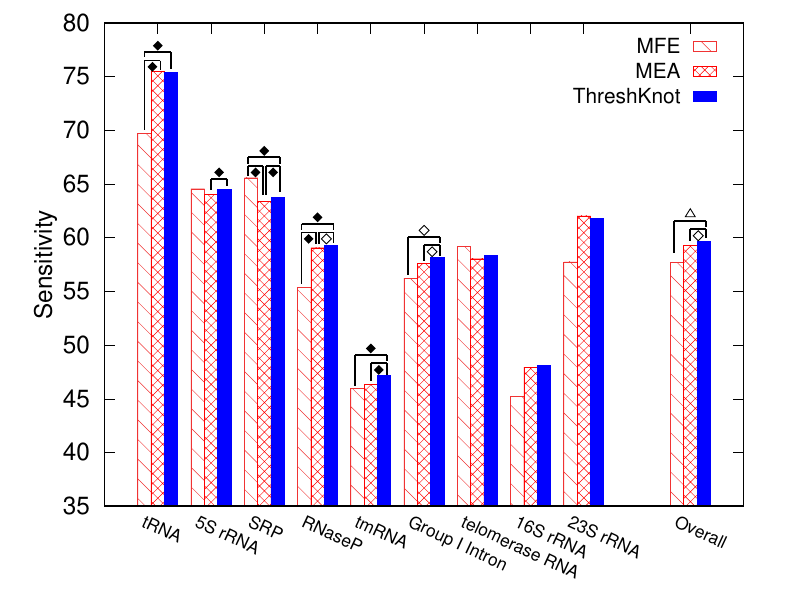}
  \end{tabular}
  \vspace{-0.3cm}
    \caption{Accuracies of MFE, MEA ($\gamma$=1.5), and \threshknot ($\theta$=0.3).
    In both panels, the first nine bars from the left represent PPV (A) and sensitivity (B) averaged over all sequences in one family, and the rightmost bars represent the overall accuracies, averaging over all families.
    Statistical significance (two-sided) is marked as $^{ \blacklozenge}$($p\! <\! 0.01$), $^{\Diamond}$($0.01 \!\leq\! p \! < \! 0.05$),  or $^{\triangle}$($0.05 \!\leq\! p \! < \! 0.06$).}
    
    \label{fig:main_text_histogram}
\end{figure*}






Figure~\ref{fig:main_text_curves} shows that the 
accuracy curve of \threshknot with varying $\theta$
is always on the upper right side of the 
accuracy curve of MEA with varying $\gamma$.
This shows that at a given level of PPV, \threshknot always has a higher sensitivity.

We further use Jackknife resampling method \cite{tukey:1958}
to choose the best parameter $\theta$ for \threshknot(see Methods) and $\gamma$ for MEA,
i.e. the parameter that maximizes the F-score (harmonic mean of sensitivity and PPV) with respect to MFE F-score. The same $\theta=0.3$ is chosen consistently across all families for \threshknot,
and the same $\gamma\!=\!1.5$ is chosen consistently for MEA,
suggesting these parameters would be widely applicable to other RNA families.
Table~\ref{tab:accuracy_table} summarizes the overall accuracies using these parameters,
comparing four methods (MFE, MEA, \probknot, and \threshknot) with \rnastructure.
\threshknot's overall sensitivity is significantly higher than MEA (+0.33\%, $p$-value 0.02)
and is the best among all methods,
while its overall PPV is only marginally and insignificantly lower than MEA (-0.02\%, $p$-value 0.97).
Figure~\ref{fig:main_text_histogram} details the accuracies on each family and 
the statistical significance tests. 

\begin{figure} [h]
  \centering
      \includegraphics[width=0.5\textwidth]{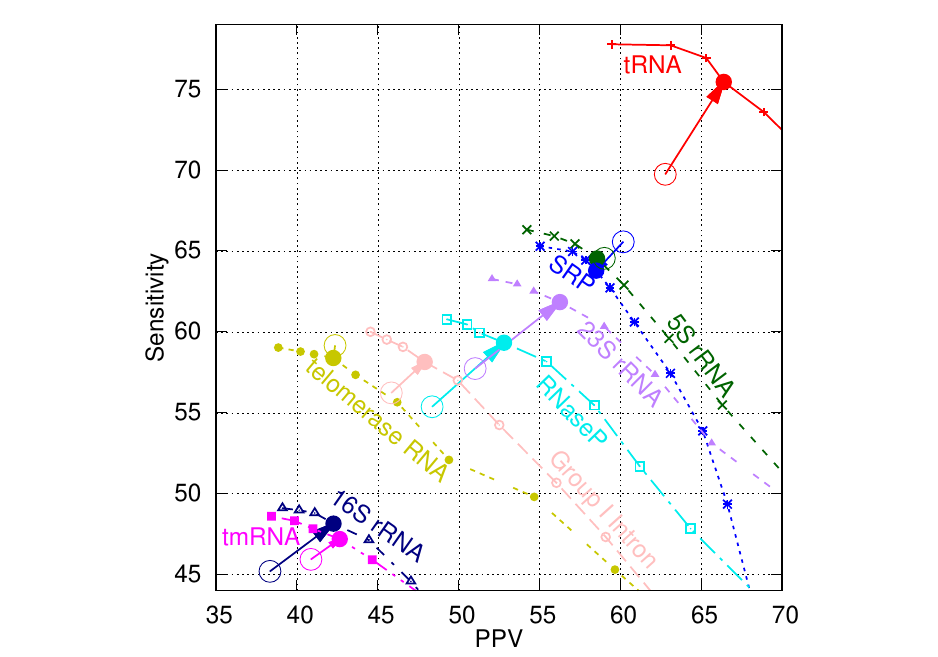}
  \caption{\threshknot improves 6 out of 9 families over MFE (in both PPV and sensitivity). 
  The curves show the \threshknot accuracies with varying $\theta$.
  The arrows point from MFE (hollow circles) to \threshknot at $\theta\!=\!0.3$.}
  \label{fig:main_text_family}
\end{figure}

\begin{figure*}[!h]
  \centering
  \begin{tabular}{ll}
{\bf A} & {\bf B}
\\
\hspace{-0.3cm} \includegraphics[width=0.4\textwidth]{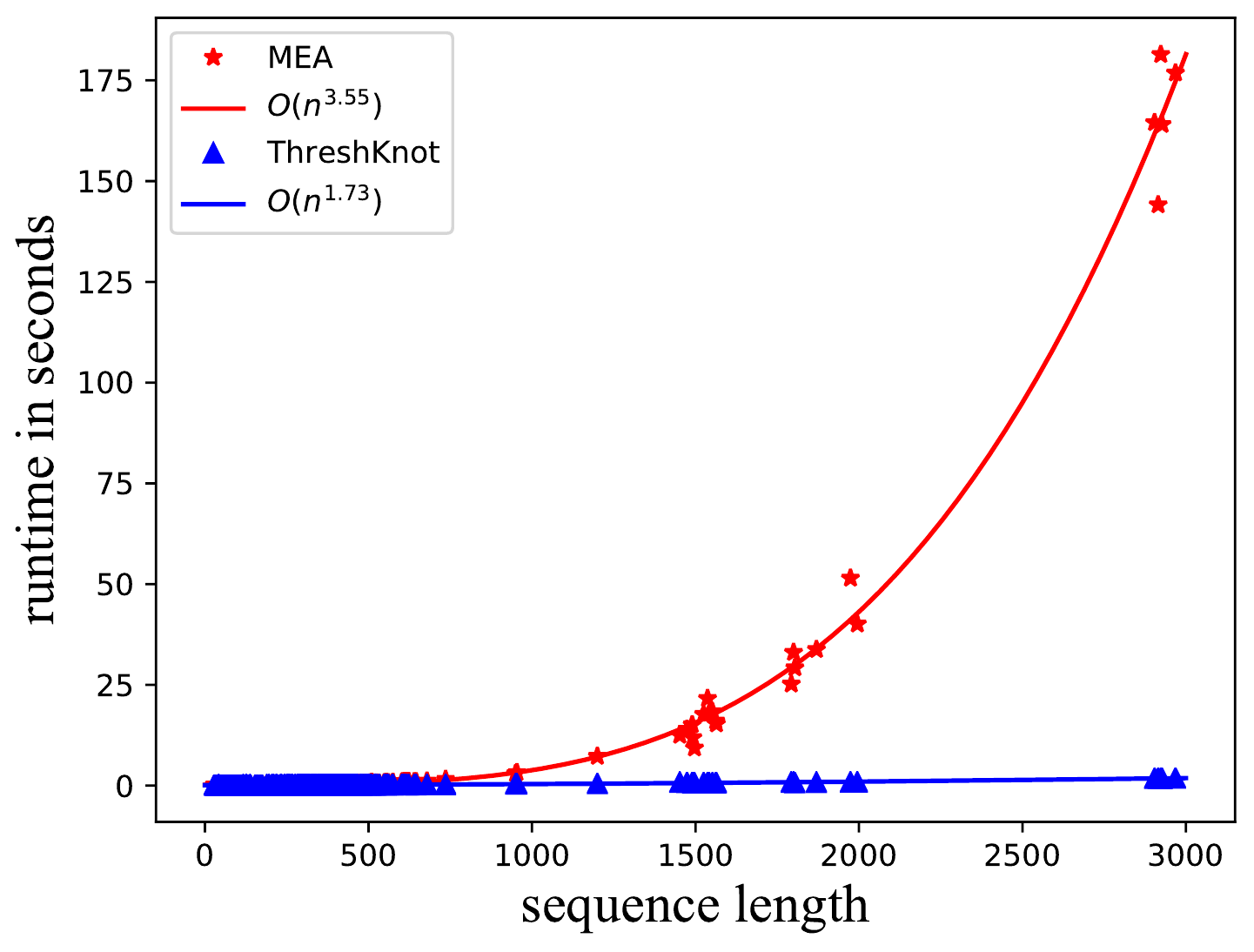}
&
\hspace{-0.2cm} \includegraphics[width=0.4\textwidth]{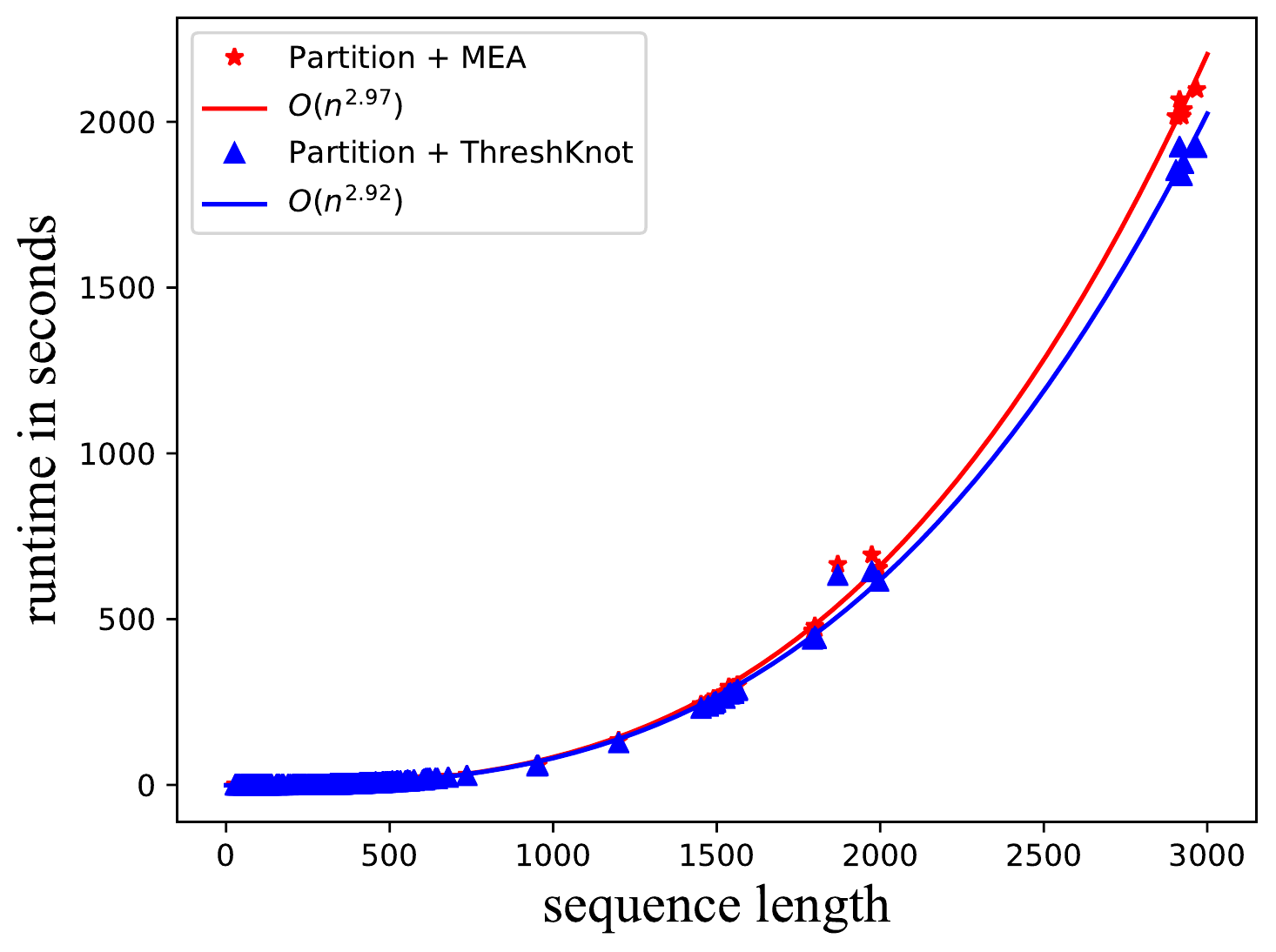}
  \end{tabular}
  \caption{Runtime comparison for the \rnastructure package on a single thread: ThreshKnot ($\theta$ = 0.3) vs.~MEA ($\gamma$ = 1.5).
  After obtaining base-pairing probabilities (the Partition part in B), MEA takes $O(n^3)$ whereas \threshknot takes $O(n^2)$ in the worst case (also see Table~\ref{tab:accuracy_table} ``time complexity'').
  {\bf A}: excluding the time for computing base-pairing probabilities (\threshknot is substantially faster than MEA).
  {\bf B}: including the time for computing base-pairing probabilities.
  }
  \label{fig:main_text_runtime}
\end{figure*}

Table~\ref{tab:accuracy_table} also includes two other systems:
\ipknot\cite{sato+:2011}
and 
\pkiss\footnote{\pkiss is the successor of pknotsRG \cite{reeder+giegerich:2004}} \cite{theis+:2010},
both of which use free energy parameters specialized for pseudoknot prediction in addition to those used by \rnastructure.
\ipknot has a higher PPV but lower sensitivity than \threshknot,
and its F-score (55.50) is slightly lower than \threshknot's (55.53);
however, it is worth noting that the \threshknot here is based on \rnastructure,
and the \threshknot versions based on \contrafold and \viennarnafold 
have higher accuracies; see Figs.~\ref{fig:Vienna_curves} and \ref{fig:CONTRAfold_original_curves}.
\pkiss, on the other hand, has lower PPV and Sensitivities.


Figure~\ref{fig:main_text_family} shows
the \threshknot accuracy curve with varying $\theta$ for each family, 
and the corresponding MFE accuracy on that family.
Compared with MFE, \threshknot improves six (6) out of nine (9) families' accuracies (in both PPV and Sensitivity).


\subsection{Pseudoknot Prediction Accuracy}

We next evaluate \threshknot's abilities to predict pseudoknots, 
and we use the PPV and sensitivity of ``crossing-pairs''
to measure the pseudoknot prediction accuracy
(see Materials and Methods for details).  
Table~\ref{tab:accuracy_table} 
compares \threshknot with \probknot, \ipknot, and \pkiss (note that MFE and MEA are unable to predict pseudoknots).
\threshknot is more accurate in pseudoknot prediction than \probknot in both crossing-pair PPV and sensitivity. 
\ipknot and \pkiss, on the other hand, are two specialized tools tailored to pseudoknot prediction,
and they indeed have higher crossing-pair PPV and sensitivity than \threshknot, which is a general-purpose structure prediction tool.
Table~\ref{tab:SI_pseudoknots_three_engines} details pseudoknot prediction accuracies for each family.



\subsection{Prediction Runtime}

We now turn to the comparison of prediction efficiency.
After obtaining base-pairing probabilities, 
\threshknot takes $O(n^2)$ time in the worst case,
whereas MEA takes $O(n^3)$ time (see Table~\ref{tab:accuracy_table} for time complexities);
this is indeed confirmed in practice by Figure~\ref{fig:main_text_runtime}A.
Furthermore, 
Fig.~\ref{fig:si_number_pij} shows that
with \threshknot, after the $O(n^2)$ threshold pruning step, 
the number of surviving base pair candidates scales linearly 
with the length of the RNA sequence
(even with a small $\theta$ such as 0.01).
This is
because the vast majority of those $O(n^2)$ pairs have close-to-zero probabilities 
(also evidenced by Figure~3B in \namecite{zuber+:2017}).
This means the core ``selection'' step of \threshknot only takes $O(n)$ time.
Therefore, as summarized in Table~\ref{tab:linear}, 
there are three steps in the whole \threshknot pipeline:
\begin{enumerate}
\item $O(n^3)$-time computation of partition function and base-pairing probabilities,
\item $O(n^2)$-time threshold pruning, and 
\item $O(n)$-time final pair selection.
\end{enumerate}
That being stated, in both \threshknot and MEA, 
the overall runtime is still dominated by the $O(n^3)$-time first step
(see Figure~\ref{fig:main_text_runtime}B). 

\begin{table}
\begin{tabular}{c|ccc}
 & base-pair & threshold & pair\\
 & probs        & pruning   & selection\\
\hline
classical (McCaskill) & $O(n^3)$ & $O(n^2)$ & $O(n)$ \\
\linearpartition & $O(n)$ & $O(n)$ & $O(n)$
\end{tabular}
\smallskip
\caption{The time complexities of \threshknot
using classical partition function calculation \cite{mccaskill:1990}
and \linearpartition \cite{zhang+:2019}.
\label{tab:linear}
}
\end{table}

\section{Discussion}
\label{sec:discussion}
In RNA secondary structure prediction, partition function-based algorithms have become increasingly popular in recent years.
Among these methods, 
MEA is popular,
but our results show that 
\threshknot always outperforms MEA  
in all three aspects: 
(1) it can achieve better overall predication accuracy, 
(2) it can predict pseudoknots that MEA can not,
(3) it is much simpler to implement and runs much faster.
This suggests that \threshknot should replace MEA as the default partition function-based structure prediction algorithm.



The overall runtime of \threshknot is still dominated by the $O(n^3)$-time first step to calculate the partition function
(i.e., the \namecite{mccaskill:1990} algorithm).
Fortunately, our forthcoming {\linearpartition} paper \cite{zhang+:2019}
reports an $O(n)$-time algorithm to approximate the partition function
inspired by the recently published \linearfold algorithm \cite{huang+:2019}, 
and it outputs just $O(n)$ base pairs with non-zero probabilities instead of all $O(n^2)$ pairs.
This implies that we can make the whole \threshknot pipeline run in $O(n)$ time
with \linearpartition (see Table~\ref{tab:linear}).


%

\vspace{-0.2cm}
\section*{Materials and Methods}


\subsection*{Dataset}

We use the ArchiveII dataset \cite{sloma+mathews:2016}, 
a diverse set of RNA sequences with accepted
structures.\footnote{\myurl{http://rna.urmc.rochester.edu/pub/archiveII.tar.gz}}
Following \linearfold \cite{huang+:2019},
we only consider full sequences (i.e., excluding the individual folding domains of 16S/23S rRNAs) and 
remove those sequences found in the S-Processed set \cite{andronescu+:2007}
(because \contrafold is trained on S-Processed).
The resulting dataset contains 2,889 sequences over 9 families, 
with an average length of 222.2 \nts and maximum length of 2,968 \nts.

\subsection*{Software and Computing Environment}

We use the following software:
\vspace{-0.1cm}
\begin{itemize}[leftmargin=10pt]
\setlength{\itemsep}{-3pt}
\item \rnastructure 6.1:\\
 \myurl{https://rna.urmc.rochester.edu/RNAstructure.html}
\item \contrafold 2.02\\ \myurl{http://contra.stanford.edu/contrafold/download.html}
\item \viennarnafold 2.4.13\\ \myurl{https://www.tbi.univie.ac.at/RNA/}
\item \ipknot\\
      \myurl{https://github.com/satoken/ipknot}
\item \pkiss\\ \myurl{https://bibiserv.cebitec.uni-bielefeld.de/pkiss}
\end{itemize}

All software were compiled by GCC 5.4.0 
on a laptop with Intel Core i7-8550U at 1.8GHz
running Ubuntu 16.04.2.

\subsection*{Evaluation Details}

We use the standard PPV and sensitivity as follows:
\[
\PPV(\yhat, \ystar) =\frac{| \yhat \cap \ystar |}{|\yhat|}, \quad \Sens(\yhat, \ystar) =\frac{| \yhat \cap \ystar |}{|\ystar|}\notag
\]
where \yhat is a predicted structure and \ystar is the accepted structure (both structures are treated as sets of pairs,
i.e., $|\yhat|$ is the number of pairs in \yhat).

Following \namecite{mathews+:1999}, 
we allow correctly predicted pairs to be offset by one position for one nucleotide as compared to the known structure (see Table~\ref{tab:SI_overall_three_engines_slipage}).
We also report in Table~\ref{tab:SI_overall_three_engines_exact}
the accuracies using exact matching.

The per-family accuracy is the mean over all sequences in that family,
and the overall accuracy is the mean over per-family accuracies from all families.

\smallskip

We use the Jackknife resampling method 
to choose the best parameter ($\theta$ for \threshknot and $\gamma$ for MEA)
as follows:
each time we held out one family, and evaluate the relative accuracy of \threshknot over MFE on
the remaining 8 families with $\theta$ ranging from 0, 0.1, 0.2, 0.3, 0.4, 0.5, and 0.6.
Coincidentally, in each case, the same $\theta=0.3$ is consistently chosen as the best paramter for \threshknot.
The same is true for $\gamma=1.5$ for MEA.
The ``relative accuracy'' is defined as the F-score between the difference in PPV and the difference in sensitivity:
\[
\Fscore(\ppv,\Sens)   = \frac{2 \cdot \ppv \cdot \Sens}{\ppv+\Sens}
\]
\begin{align}
 \Delta \Fscore\big((\ppv',\Sens'), (\ppv, \Sens)\big)    
 =                   \Fscore(&\ppv'-\ppv, \notag\\
               & \Sens'-\Sens)\notag
 \end{align}
Where $(\ppv',\Sens')$ are the PPV and sensitivity of \threshknot and $(\ppv, \Sens)$
are those of MFE
(we assume $\ppv'>\ppv$ and $\Sens'>\Sens$).


\smallskip

For pseudoknot accuracy, we first define the notion of ``crossing pairs'', notated $\crossing(y)$,
in a structure $y$ to be the set of pairs
that are crossed by at least one other pair:
\begin{align}
  \crossing(y) =&   \{ (i,j) \in y \mid  \exists (k,l) \in y, \notag\\
  & \qquad i\!<\! k \!<\!j\!<\!l\text{ or } k\!<\!i\!<\!l\!<\!j \} \notag
\end{align}

We then restrict ourselves to comparing the crossing pairs in the predicted structure
to the crossing pairs in the accepted structure,
and define the pseudoknot PPV and sensitivity to be the PPV and sensitivity on those two subsets:
\begin{align}
\nonumber
\ppvcrossing(\yhat, \ystar) &=\PPV(\crossing(\yhat), \crossing(\ystar)) \notag\\
\senscrossing(\yhat, \ystar) &=\Sens(\crossing(\yhat), \crossing(\ystar)) \notag
\end{align}
This means that a crossing pair in the predicted structure $\yhat$
is considered correct if it is also a crossing pair in the 
accepted structure $\ystar$.


%
\smallskip

All statistical significance tests are done with two-sided permutation test \cite{aghaeepour+hoos:2013}.

\subsection*{Code Availability}

ThreshKnot is available in the \rnastructure software package v6.2 (released November 27, 2019):
 \myurl{https://rna.urmc.rochester.edu/RNAstructure.html}
 
To run \threshknot in the \rnastructure package:

\noindent{\scriptsize \verb|./ProbKnot --sequence <infile> <outfile> -t 0.3 -m 2|}
 
 Where \verb|-t| specifies a threshold probability to include a pair; \verb|-m| specifies the minimum length accepted for a helix.
 We set threshold $\theta=0.3$ and the minimum helix length as 2 for ThreshKnot using \rnastructure.

 \subsection*{Data Availability}
 The data that support our findings  are available from the corresponding author upon request.

\acknow{
This work was partially supported by 
NSF grant IIS-1817231 (L.H.) and NIH grant R01 GM076485 (D.H.M.).
}





\showacknow{} 
\smallskip


\balance

\bibliographystyle{elsarticle-harv}

\bibliography{main}

\appendix

\newpage

\newpage
\onecolumn
  \begin{centering}
          \textbf{\large Supporting Information}\\
    \vspace{0.5cm}
    \textbf{\large ThreshKnot: Thresholded ProbKnot for Improved RNA Secondary Structure Prediction}\\
    \vspace{0.5cm}
    \textbf{
     \small Liang Zhang, He Zhang, David H.~Mathews, and Liang Huang}
    
  \end{centering}

\setcounter{figure}{0}
\renewcommand{\thefigure}{SI\,\arabic{figure}} 

\setcounter{table}{0}
\renewcommand{\thetable}{SI\,\arabic{table}}




\begin{figure*}[h]
\centering
\begin{minipage}{.45\linewidth}
  \includegraphics[width=\linewidth]{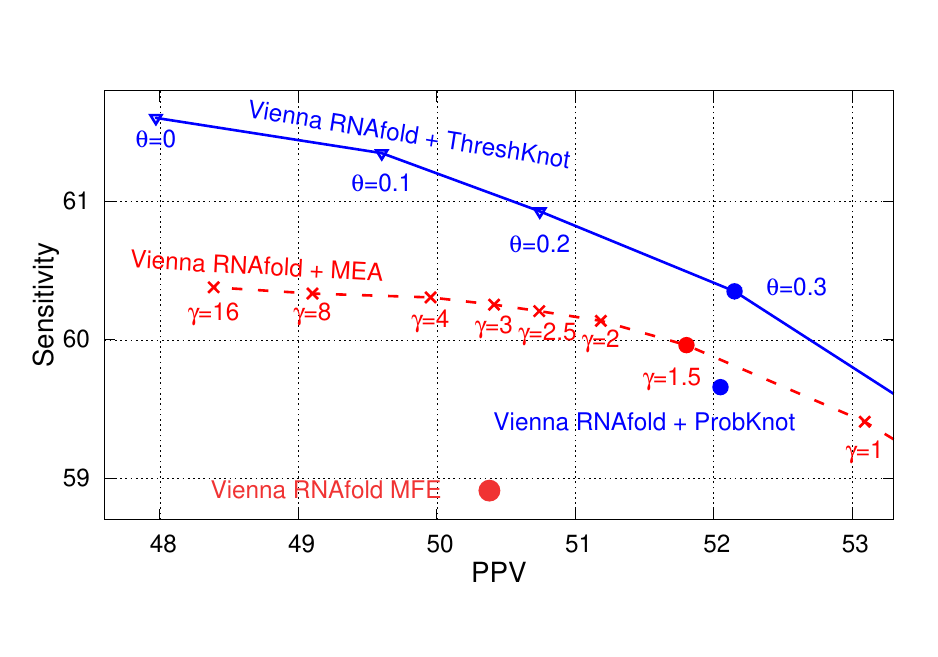}
  \captionof{figure}{Comparison between \threshknot, \probknot, MFE, and MEA on \viennarna using the ArchiveII dataset.
  \threshknot has better PPV and Sensitivity than all other methods.
  We also add \probknot for comparison.}
  \label{fig:Vienna_curves}
\end{minipage}
\hspace{.05\linewidth}
\begin{minipage}{.45\linewidth}
  \includegraphics[width=\linewidth]{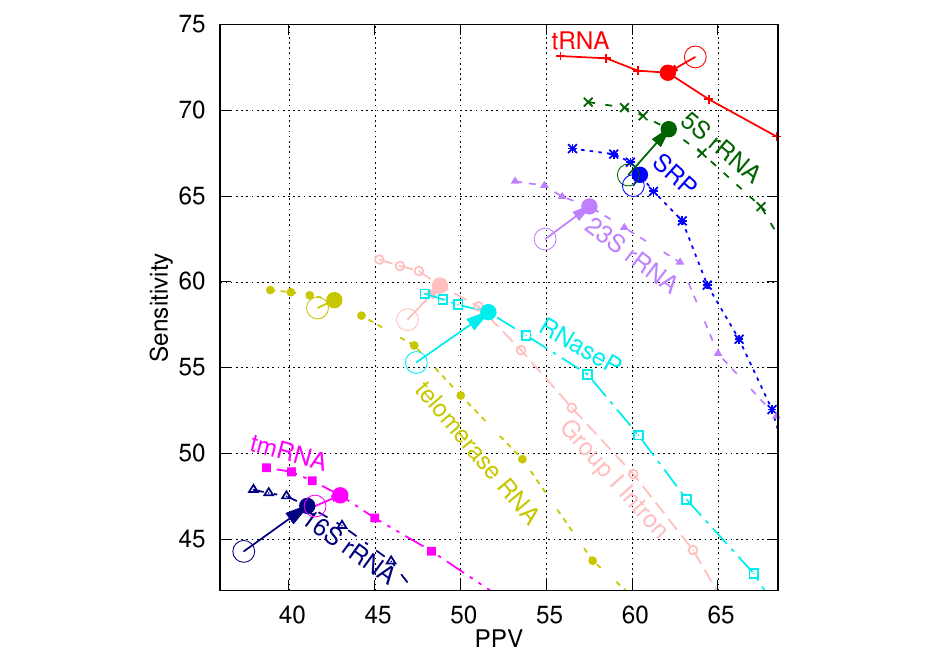}
  \captionof{figure}{\threshknot improves 8 out of 9 families over MFE (in both PPV and sensitivity) on \viennarnafold. 
  The curves show the \threshknot accuracies with varying $\theta$.
  The arrows point from MFE (hollow circles) to \threshknot at $\theta\!=\!0.3$.}
  \label{fig:Vienna_families}
\end{minipage}
\end{figure*}

\begin{figure}
\centering
\begin{minipage}{.45\linewidth}
  \includegraphics[width=\linewidth]{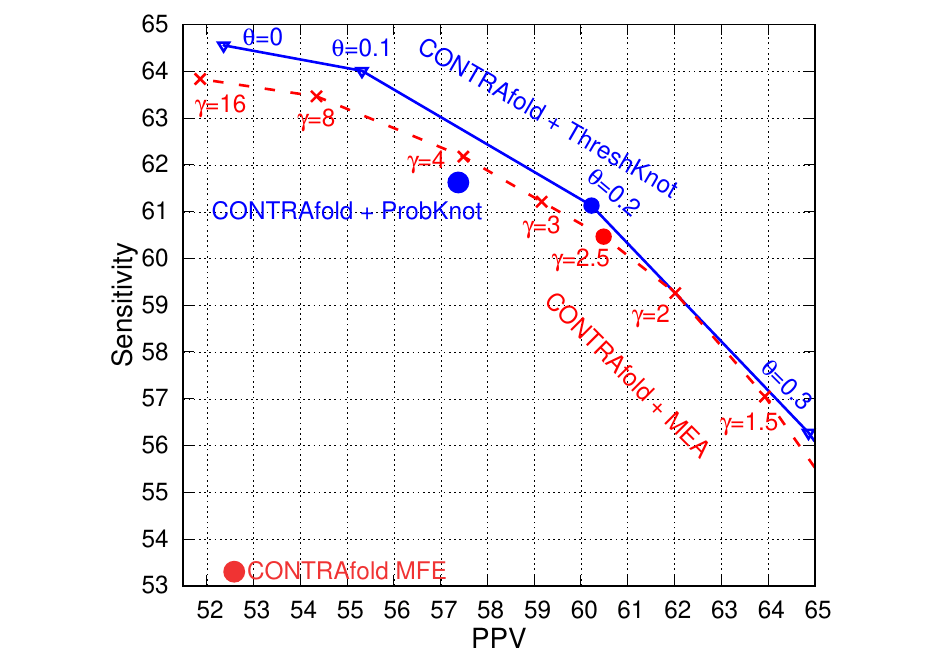}
  \captionof{figure}{Comparison between \threshknot, \probknot, MFE, and MEA on \contrafold using the ArchiveII dataset.
  \threshknot has better PPV and Sensitivity than all other methods.}
  \label{fig:CONTRAfold_original_curves}
\end{minipage}
\hspace{.05\linewidth}
\begin{minipage}{.45\linewidth}
  \includegraphics[width=\linewidth]{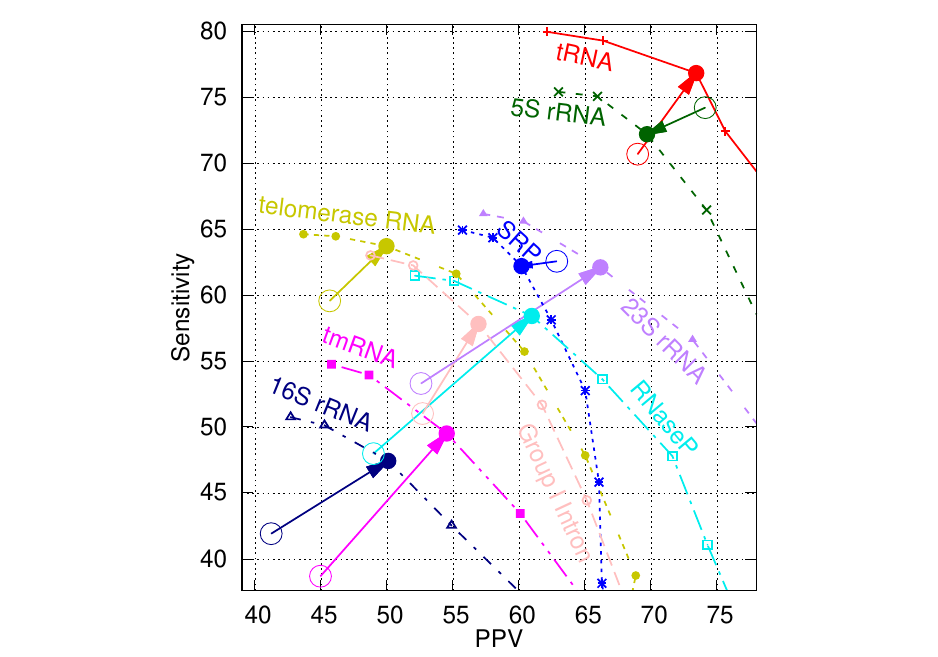}
  \captionof{figure}{\threshknot improves 7 out of 9 families over MFE (in both PPV and sensitivity) on \contrafold. 
  The curves show the \threshknot accuracies with varying $\theta$.
  The arrows point from MFE (hollow circles) to \threshknot at $\theta\!=\!0.2$.}
  \label{fig:CONTRAfold_original_families}
\end{minipage}
\end{figure}

\begin{figure*}
\centering
  \begin{tabular}{ll}
\quad {\bf A} & \quad {\bf B}
\\
\hspace{-0.7cm}
    \includegraphics[width=0.53\textwidth]{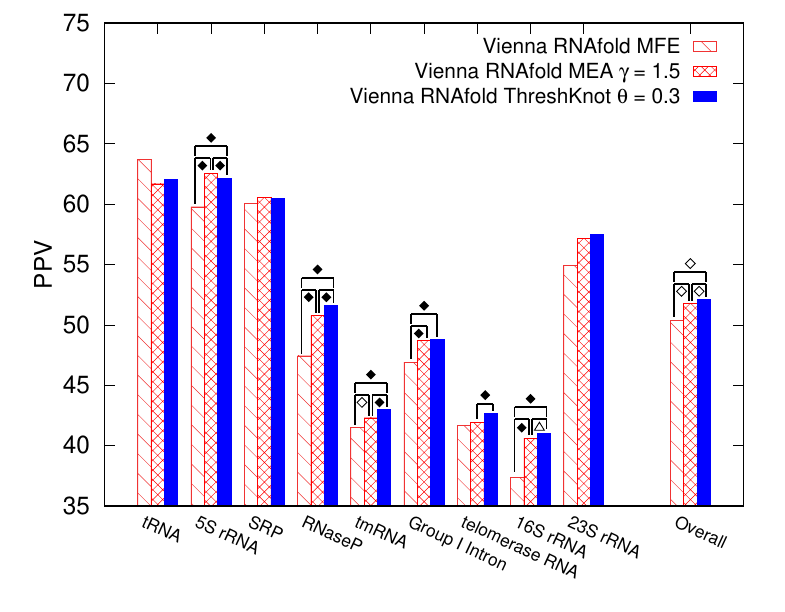}
&
\hspace{-0.7cm}
    \includegraphics[width=0.53\textwidth]{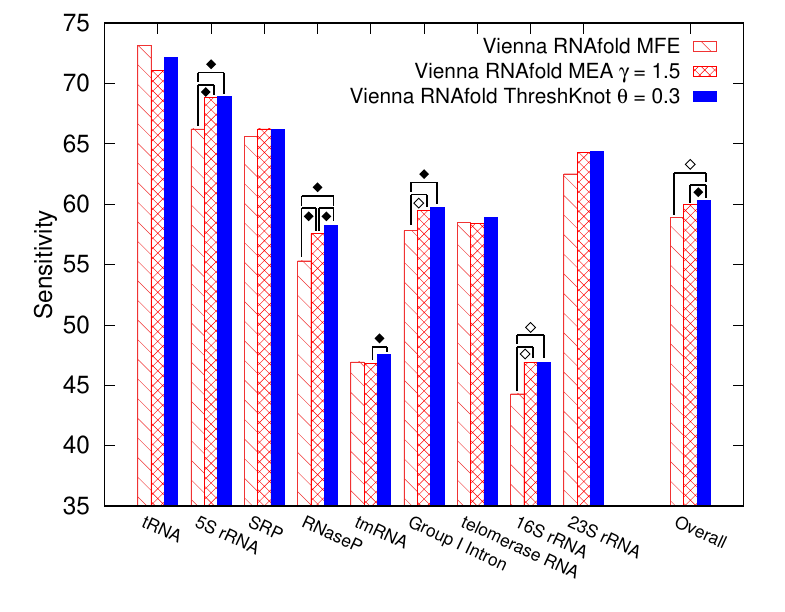}
\end{tabular}
    \caption{Accuracy results of MFE, MEA, and \threshknot using \viennarnafold.
In both panels, the first nine bars from the left represent PPV (A) and sensitivity (B) averaged over all sequences in one family, and the rightmost bars represent the overall accuracies, averaging over all families.
    Statistical significance (two-sided) is marked as $^{ \blacklozenge}$($p\! <\! 0.01$), $^{\Diamond}$($0.01 \!\leq\! p \! < \! 0.05$),  or $^{\triangle}$($0.05 \!\leq\! p \! < \! 0.06$).}
    
    \label{fig:SI_vienna_histogram}
\end{figure*}

\begin{figure*}
\centering
  \begin{tabular}{ll}
\quad {\bf A} & \quad {\bf B}
\\
\hspace{-0.7cm}
    \includegraphics[width=0.53\textwidth]{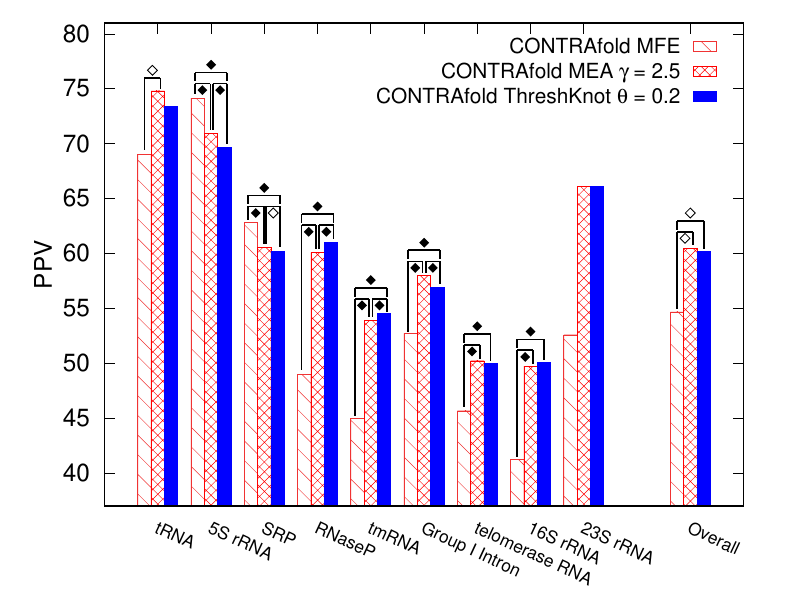}
&
\hspace{-0.7cm}
    \includegraphics[width=0.53\textwidth]{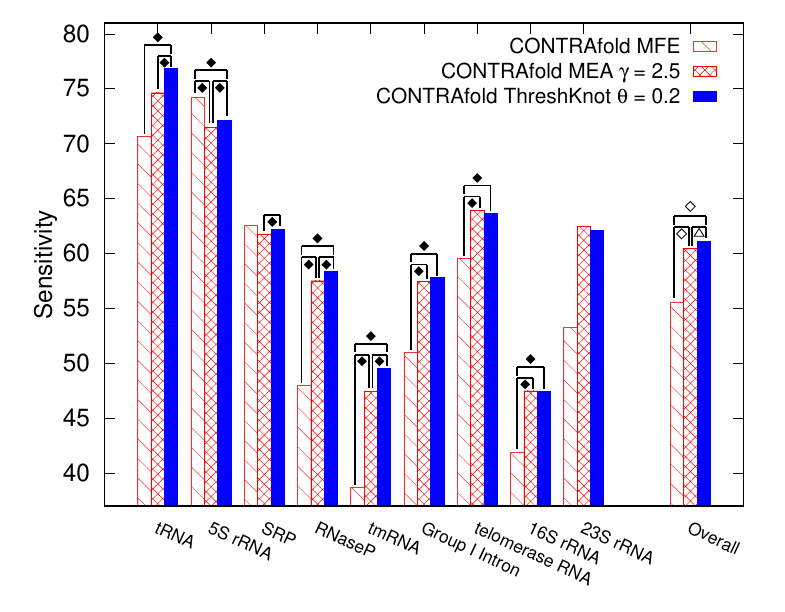}
\end{tabular}
    \caption{Accuracy results of MFE, MEA, and \threshknot using \contrafold.
In both panels, the first nine bars from the left represent PPV (A) and sensitivity (B) averaged over all sequences in one family, and the rightmost bars represent the overall accuracies, averaging over all families.
    Statistical significance (two-sided) is marked as $^{ \blacklozenge}$($p\! <\! 0.01$), $^{\Diamond}$($0.01 \!\leq\! p \! < \! 0.05$),  or $^{\triangle}$($0.05 \!\leq\! p \! < \! 0.06$).}
    
    \label{fig:SI_contrafold_original_histogram}
\end{figure*}

\begin{figure} 
  \centering
      \includegraphics[width=0.6\textwidth]{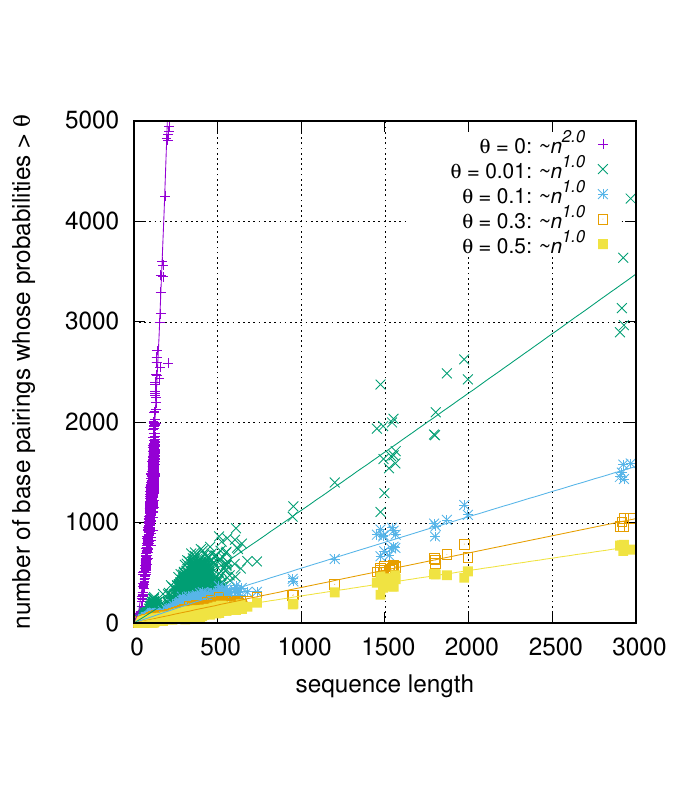}
  \caption{The number of base pairs whose probabilities are greater than the threshlds $\theta$
  in the base pairing probability matrix from \rnastructure.
  With any non-zero $\theta$, the number of surviving base pairs is linear in sequence length. 
  }
  \label{fig:si_number_pij}
\end{figure}

\newpage

\begin{table}[]
\centering
\setlength{\tabcolsep}{1.mm}{
\resizebox{1\textwidth}{!}{
{\tabcolsep=5pt
\begin{tabular}[b]{rrc|b{0.06\textwidth}<{\raggedbottom}b{0.06\textwidth}<{\raggedbottom}b{0.06\textwidth}<{\raggedbottom}b{0.06\textwidth}<{\raggedbottom}b{0.06\textwidth}<{\raggedbottom}b{0.06\textwidth}<{\raggedbottom}b{0.06\textwidth}<{\raggedbottom}b{0.06\textwidth}<{\raggedbottom}b{0.06\textwidth}<{\raggedbottom}|r}
                       &                             \multicolumn{2}{r|}{ Family} 			& tRNA  & 5S rRNA & SRP   & RNaseP & tmRNA & Group I Intron & telomerase RNA & 16S rRNA & 23S rRNA & Overall \\
\hline
		&	\multicolumn{2}{r|}{      total seqs}		 & 557   & 1,283   & 928   & 454    & 462   & 98             & 37             & 22       &  5        & 3,846   \\
		& 	\multicolumn{2}{r|}{      used seqs}		 & 74    & 1,125   & 886   & 182    & 462   & 96             & 37             & 22       & 5        & 2,889   \\
		&\multicolumn{2}{r|}{avg. length of used seqs}    & 77.3  & 118.8   & 186.1 & 344.1  & 366.0 & 424.9          & 444.6          & 1,547.9  & 2,927.4  & 222.2   \\
\hline
\hline
\multirow{6}{*}{\rnastructure}          	& \multirow{2}{*}{MFE}					& PPV    & 62.77 & 59.01   & 60.18 & 48.36  & 40.87 & 45.84          & 42.37          & 38.34    & 51.02    & 49.86   \\
                      					&                             						& sens   & 69.73 & 64.54   & 65.56 & 55.36  & 45.93 & 56.22          & 59.15          & 45.19    & 57.73    & 57.71   \\
                      					\cline{2-13}
                      					& \multirow{1}{*}{MEA} & PPV    & 68.11 & 58.73   & 58.41 & 52.42  & 42.17 & 47.87          & 41.86          & 41.98    & 56.25    & 51.98   \\
                      					& \multirow{1}{*}{ $\gamma=1.5$}  & sens   & 75.45 & 64.01   & 63.35 & 59.05  & 46.35 & 57.59          & 58.03          & 47.96    & 62.01    & 59.31   \\
							\cline{2-13}
                      					& \multirow{1}{*}{ThreshKnot} 	& PPV    & 66.39 & 58.56   & 58.50 & 52.80  & 42.65 & 47.91          & 42.25          & 42.27    & 56.27    & 51.96   \\
                      					& \multirow{1}{*}{$\theta=0.3$} 	& sens   & 75.44 & 64.53   & 63.78 & 59.33  & 47.19 & 58.14          & 58.38          & 48.14    & 61.84    & 59.64   \\
\hline
\hline                      
\multirow{6}{*}{\rnafold}        	& \multirow{2}{*}{MFE}		                         	& PPV    & 63.69 & 59.79   & 60.08 & 47.43  & 41.53 & 46.91          & 41.67          & 37.37    & 54.94    & 50.38   \\
						     	&		         						& sens   & 73.11 & 66.22   & 65.61 & 55.30  & 46.93 & 57.80          & 58.48          & 44.29    & 62.49    & 58.91   \\
							\cline{2-13}
                      					& \multirow{1}{*}{MEA}			               & PPV    & 61.68 & 62.52   & 60.56 & 50.76  & 42.30 & 48.70          & 41.91          & 40.61    & 57.19    & 51.80   \\
                      					&\multirow{1}{*}{$\gamma=1.5$} 			& sens   & 71.09 & 68.80   & 66.22 & 57.57  & 46.85 & 59.49          & 58.43          & 46.90    & 64.29    & 59.96   \\
							\cline{2-13}
                      					& \multirow{1}{*}{ThreshKnot} 	& PPV    & 62.09 & 62.14   & 60.46 & 51.63  & 43.00 & 48.81          & 42.65          & 41.07    & 57.51    & 52.15   \\
                      					& \multirow{1}{*}{$\theta=0.3$} 	& sens   & 72.18 & 68.90   & 66.24 & 58.23  & 47.56 & 59.78          & 58.93          & 46.94    & 64.39    & 60.35   \\
\hline
\hline                      
\multirow{6}{*}{\contrafold} 		& \multirow{2}{*}{MFE}		                         & PPV    & 69.00 & 74.12   & 62.87 & 48.99  & 44.97 & 52.71          & 45.67          & 41.23    & 52.59    & 54.68   \\
                      					&                             						& sens   & 70.67 & 74.20   & 62.55 & 47.98  & 38.69 & 51.01          & 59.56          & 41.92    & 53.30    & 55.54   \\
							\cline{2-13}
                      					& \multirow{1}{*}{MEA} 					& PPV   & 74.80 & 70.96   & 60.58 & 60.09  & 53.89 & 58.00          & 50.23          & 49.70    & 66.13    & 60.49   \\
                      					& \multirow{1}{*}{$\gamma=2.5$}              		& sens   & 74.63 & 71.52   & 61.71 & 57.51  & 47.46 & 57.48          & 63.95          & 47.49    & 62.45    & 60.47   \\
							\cline{2-13}
                      					& \multirow{1}{*}{ThreshKnot}				& PPV    & 73.43 & 69.71   & 60.20 & 60.98  & 54.53 & 56.94          & 49.97          & 50.11    & 66.17    & 60.23   \\
                      					& \multirow{1}{*}{$\theta=0.2$}				& sens   & 76.82 & 72.18   & 62.19 & 58.41  & 49.51 & 57.80          & 63.70          & 47.43    & 62.10    & 61.13   \\
\hline
\hline                      
\multirow{2}{*}{IPknot}	                &						                         & PPV    & 81.89 & 62.53   & 56.63 & 65.24  & 55.71 & 55.98          & 43.00          & 52.96    & 68.07    & 60.22   \\
                      					&                             						& sens   & 80.25 & 50.12   & 49.07 & 56.89  & 43.15 & 48.80          & 44.44          & 41.45    & 48.98    & 51.46   \\
\hline
\hline                      
\multirow{2}{*}{pKiss}	                 &						                         & PPV    & 47.82 & 47.19   & 54.45 & 41.65  & 36.70 & 47.09          & 38.58          & 38.63    & 46.74    & 44.32   \\
                      					&                             						& sens   & 55.16 & 50.45   & 59.03 & 46.80  & 40.93 & 57.70          & 53.38          & 44.48    & 51.32    & 51.03  \\

\end{tabular}
}
}
\caption{Detailed overall prediction accuracies, allowing one nucleotide in a pair to be displaced by one position, on the ArchiveII dataset.
This slipping method~\cite{mathews+:1999}  considers a base pair to be correct if it is slipped by one nucleotide on a strand.
}
\label{tab:SI_overall_three_engines_slipage}}
\end{table}

\begin{table}[]
\centering
\setlength{\tabcolsep}{1.mm}{
\resizebox{1\textwidth}{!}{
{\tabcolsep=5pt
\begin{tabular}{rrc|b{0.06\textwidth}<{\raggedbottom}b{0.06\textwidth}<{\raggedbottom}b{0.06\textwidth}<{\raggedbottom}b{0.06\textwidth}<{\raggedbottom}b{0.06\textwidth}<{\raggedbottom}b{0.06\textwidth}<{\raggedbottom}b{0.06\textwidth}<{\raggedbottom}b{0.06\textwidth}<{\raggedbottom}b{0.06\textwidth}<{\raggedbottom}|r}
                       &                             \multicolumn{2}{r|}{ Family} 			& tRNA  & 5S rRNA & SRP   & RNaseP & tmRNA & Group I Intron & telomerase RNA & 16S rRNA & 23S rRNA & Overall \\
\hline
		&	\multicolumn{2}{r|}{      total seqs}		 & 557   & 1,283   & 928   & 454    & 462   & 98             & 37             & 22       &  5        & 3,846   \\
		& 	\multicolumn{2}{r|}{      used seqs}		 & 74    & 1,125   & 886   & 182    & 462   & 96             & 37             & 22       & 5        & 2,889   \\
		&\multicolumn{2}{r|}{avg. length of used seqs}    & 77.3  & 118.8   & 186.1 & 344.1  & 366.0 & 424.9          & 444.6          & 1,547.9  & 2,927.4  & 222.2   \\
\hline
\hline
\multirow{6}{*}{\rnastructure}          	& \multirow{2}{*}{MFE}            		            	& PPV    & 61.49 & 56.55   & 56.84 & 46.46  & 38.65 & 44.13          & 39.99          & 36.52    & 48.86    & 47.72   \\
                      					&                            						& sens   & 68.39 & 61.77   & 61.67 & 53.08  & 43.41 & 54.07          & 55.79          & 43.05    & 55.28    & 55.17   \\
                                            		\cline{2-13}
                      					& \multirow{1}{*}{MEA}			               & PPV    & 65.95 & 56.36   & 55.17 & 50.40  & 39.58 & 46.23          & 39.45          & 40.43    & 54.31    & 49.76   \\
                      					& \multirow{1}{*}{$\gamma=1.5$}      		         & sens   & 73.01 & 61.34   & 59.67 & 56.74  & 43.50 & 55.64          & 54.63          & 46.17    & 59.88    & 56.73   \\
                      					\cline{2-13}
                      					& \multirow{1}{*}{ThreshKnot} 	& PPV    & 64.15 & 56.25   & 55.28 & 50.83  & 40.06 & 46.25          & 39.90          & 40.70    & 54.39    & 49.76   \\
                      					&  \multirow{1}{*}{$\theta=0.3$} 	& sens   & 72.87 & 61.89   & 60.10 & 57.07  & 44.32 & 56.12          & 55.07          & 46.35    & 59.78    & 57.06   \\
\hline
\hline
\multirow{6}{*}{\rnafold}        	& \multirow{2}{*}{MFE}		                         	& PPV    & 61.75 & 57.28   & 56.58 & 45.76  & 39.75 & 45.49          & 39.53          & 35.65    & 53.20    & 48.33   \\
						     	&		         						& sens   & 70.98 & 63.35   & 61.55 & 53.28  & 44.90 & 56.06          & 55.40          & 42.26    & 60.50    & 56.48   \\
							\cline{2-13}
                      					& \multirow{1}{*}{MEA}               & PPV    & 59.72 & 60.08   & 57.13 & 48.98  & 40.53 & 47.13          & 39.42          & 38.84    & 55.54    & 49.71   \\
                      					& \multirow{1}{*}{$\gamma=1.5$}       	& sens   & 68.89 & 66.01   & 62.22 & 55.48  & 44.88 & 57.60          & 54.89          & 44.85    & 62.43    & 57.47   \\
							\cline{2-13}
                      					& \multirow{1}{*}{ThreshKnot} 	& PPV    & 60.20 & 59.73   & 57.07 & 49.87  & 41.20 & 47.28          & 40.20          & 39.38    & 55.80    & 50.08   \\
                      					&  \multirow{1}{*}{$\theta=0.3$} 	& sens   & 70.04 & 66.12   & 62.30 & 56.19  & 45.55 & 57.92          & 55.50          & 45.01    & 62.46    & 57.90   \\
\hline
\hline
\multirow{6}{*}{\contrafold} 		& \multirow{2}{*}{MFE}		                         & PPV    & 67.61 & 70.68   & 59.14 & 47.45  & 42.96 & 51.21          & 43.40          & 39.84    & 50.56    & 52.54   \\
                      					&                             						& sens   & 69.12 & 70.70   & 58.61 & 46.39  & 36.94 & 49.56          & 56.58          & 40.49    & 51.24    & 53.29   \\
							\cline{2-13}
                      					& \multirow{1}{*}{MEA}         		     	& PPV    & 73.56 & 67.94   & 57.06 & 58.26  & 51.68 & 56.43          & 47.45          & 48.09    & 64.15    & 58.29   \\
                      					& \multirow{1}{*}{$\gamma=2.5$}              	& sens   & 73.27 & 68.31   & 57.94 & 55.62  & 45.50 & 55.94          & 60.37          & 45.95    & 60.56    & 58.16   \\
							\cline{2-13}
                      					& \multirow{1}{*}{ThreshKnot}	& PPV    & 72.19 & 67.01   & 56.85 & 59.17  & 52.36 & 55.49          & 47.28          & 48.61    & 64.28    & 58.14   \\
                      					& \multirow{1}{*}{$\theta=0.2$}	& sens   & 75.47 & 69.24   & 58.56 & 56.53  & 47.53 & 56.36          & 60.21          & 46.01    & 60.30    & 58.91   \\

\hline
\hline                      
\multirow{2}{*}{IPknot}	                &						                         & PPV    & 80.28 & 59.65   & 54.03 & 63.62  & 53.93 & 54.41          & 40.92          & 51.78    & 66.28    & 58.33   \\
                      					&                             						& sens   & 78.51 & 47.66   & 46.66 & 55.37  & 41.73 & 47.48          & 42.25          & 40.51    & 47.67    & 49.76   \\
\hline
\hline                      
\multirow{2}{*}{pKiss}	                 &						                         & PPV    & 45.90 & 45.14   & 51.04 & 40.19  & 34.56 & 45.62          & 37.11          & 37.21    & 44.88    & 42.41   \\
                      					&                             						& sens   & 53.04 & 48.17   & 55.13 & 45.09  & 38.54 & 55.90          & 51.29          & 42.83    & 49.27    & 48.80  \\

\end{tabular}
}
}
\caption{Detailed overall prediction accuracies on the ArchiveII dataset. The accuracies use exact base-pair matching.
}
\label{tab:SI_overall_three_engines_exact}}
\end{table}

\begin{table}[]
\centering
\setlength{\tabcolsep}{1.mm}{
\resizebox{1\textwidth}{!}{
{\tabcolsep=5pt
\begin{tabular}[b]{rr|b{0.06\textwidth}<{\raggedbottom}b{0.06\textwidth}<{\raggedbottom}b{0.06\textwidth}<{\raggedbottom}b{0.06\textwidth}<{\raggedbottom}b{0.06\textwidth}<{\raggedbottom}b{0.06\textwidth}<{\raggedbottom}b{0.06\textwidth}<{\raggedbottom}b{0.06\textwidth}<{\raggedbottom}b{0.06\textwidth}<{\raggedbottom}|r}
&						family                               & tRNA   & 5S rRNA & SRP    & RNaseP & tmRNA   & Group I Intron & telomerase RNA & 16S rRNA & 23S rRNA & Overall\\
\hline
&         gold base pairs					             & 1,496   & 37,727 & 49,680 & 17,308  & 45,332         & 9,669          & 3,774    & 9,135    & 4,091   & 178,212 \\
 &        gold crossing pairs						       & 0       & 0      & 0      & 4,538   & 26,153         & 1,164          & 1,015    & 568      & 443     & 33,881  \\
\hline
\hline
								& predicted base pairs        & 1,734   & 41,755 & 54,455 & 19,527  & 50,153         & 12,433         & 5,278    & 10,699   & 4,498   & 200,532 \\
                \rnastructure 				& predicted crossing pairs  & 167     & 2,064  & 3,218  & 1,254   & 4,510          & 929            & 407      & 880      & 445     & 13,874  \\
		+ \threshknot				& correct crossing pairs& 0       & 0      & 0      & 139     & 983            & 48             & 13       & 6        & 17      & 1,206   \\
\cline{2-12}
                 $\theta=0.3$              		& PPV                                  & 0       & 0      & 0      & 11.08 & 21.80        & 5.17         & 3.19   & 0.68   & 3.82  & 7.62  \\
                                     				& sens                                 & NA      & NA     & NA     & 3.06  & 3.76         & 4.12         & 1.28   & 1.06   & 3.84  & 2.85  \\
\hline
\hline
								     	& predicted base pairs        & 1,776   & 41,986 & 54,907 & 19,575  & 50,054         & 12,522         & 5,272    & 10,653   & 4,583   & 201,328 \\
                \rnafold					& predicted crossing pairs  & 183     & 2,867  & 2,716  & 1,263   & 4,882          & 945            & 276      & 1,013    & 325     & 14,470  \\
		+ \threshknot					& correct crossing pairs& 0       & 0      & 0      & 185     & 965            & 61             & 5        & 22       & 5       & 1,243   \\
\cline{2-12}
                $\theta=0.3$				& PPV                                  & 0       & 0      & 0      & 14.65 & 19.77        & 6.46         & 1.81   & 2.17   & 1.54  & 7.73  \\
                                     				& sens                                 & NA      & NA     & NA     & 4.08  & 3.69         & 5.24         & 0.49   & 3.87   & 1.13  & 3.08  \\
\hline
\hline
									& predicted base pairs        & 1,610   & 38,798 & 50,296 & 16,756  & 40,939         & 10,266         & 4,808    & 8,901    & 3,837   & 176,211 \\
                  \contrafold       					& predicted crossing pairs  & 319     & 4,998  & 5,684  & 1,912   & 7,893          & 1,140          & 650      & 1,263    & 561     & 24,420  \\
                 + \threshknot  					& correct crossing pairs& 0       & 0      & 0      & 307     & 2,741          & 111            & 48       & 13       & 18      & 3,238   \\
\cline{2-12}
                    $\theta=0.2$		                 & PPV                                  & 0       & 0      & 0      & 16.06 & 34.73        & 9.74         & 7.38   & 1.03   & 3.21  & 12.02 \\
                                     				& sens                                 & NA      & NA     & NA     & 6.77  & 10.48        & 9.54         & 4.73   & 2.29   & 4.06  & 6.31  \\
\hline
\hline
\multirow{5}{*}{\ipknot}			         & predicted base pairs        & 1,494   & 30,680 & 41,545 & 15,165  & 34,982         & 8,745          & 3,874    & 7,256    & 2,947   & 146,688 \\
                                     				& predicted crossing pairs  & 140     & 3,664  & 5,499  & 1,770   & 6,407          & 1,096          & 712      & 982      & 358     & 20,628  \\
                                     				& correct crossing pairs& 0       & 0      & 0      & 470     & 2,155          & 78             & 75       & 37       & 55      & 2,870   \\
\cline{2-12}
                                     & PPV                                  & 0       & 0      & 0      & 26.55 & 33.64        & 7.12         & 10.53  & 3.77   & 15.36 & 16.16 \\
                                     & sens                                 & NA      & NA     & NA     & 10.36 & 8.24         & 6.70         & 7.39   & 6.51   & 12.42 & 8.60  \\
\hline
\hline
\multirow{5}{*}{\pkiss}		                 & predicted base pairs        & 1,755   & 40,106 & 54,149 & 19,596  & 50,505         & 12,598         & 5,301    & 10,766   & 4,486   & 199,262 \\
                                     				& predicted crossing pairs  & 516     & 8,686  & 10,284 & 5,309   & 21,728         & 4,199          & 1,374    & 3,009    & 1,416   & 56,521  \\
                                     				& correct crossing pairs& 0       & 0      & 0      & 356     & 7,273          & 289            & 74       & 76       & 47      & 8,115   \\
\cline{2-12}
                                     & PPV                                  & 0       & 0      & 0      & 6.71  & 33.47        & 6.88         & 5.39   & 2.53   & 3.32  & 9.72  \\
                                     & sens                                 & NA      & NA     & NA     & 7.84  & 27.81        & 24.83        & 7.29   & 13.38  & 10.61 & 15.29\\
\end{tabular}
}
}
\caption{Detailed pseudoknots prediction accuracies, allowing one nucleotide in a pair to be displaced by one position, on the ArchiveII dataset.
This slipping method considers a base pair to be correct if it is slipped by one nucleotide on a strand.
For pseudoknot prediction accuracy, 
we compare all crossing pairs 
in the predicted structure $\yhat$
with all crossing pairs in the accepted structure $\ystar$.
A crossing pair in predicted structure $\yhat$
is considered correct if it is also a crossing pair in the 
accepted structure $\ystar$.
}
\label{tab:SI_pseudoknots_three_engines}}
\end{table}

\end{document}